\documentclass[structabstract]{aa} 
\usepackage{natbib}
\usepackage{graphicx}
\usepackage{epsf}
\usepackage{txfonts}
\bibpunct{(}{)}{;}{a}{}{,}

\newcommand{\kms}{km\,s$^{-1}$}
\newcommand{\fuse}{\emph{FUSE}}
\newcommand{\uves}{\emph{UVES}}

\newcommand{\iue}{\emph{IUE}}

\newcommand{\vsini}{$v$\,sin\,$i$}

\newcommand{\teff}{$T_{\rm eff}$}
\newcommand{\logg}{log \it g\rm~}

\newcommand{\msol}{$M_{\odot}$}
\newcommand{\lsol}{$L_{\odot}$}

\newcommand{\synspec}{{\sc Synspec}}

\newcommand{\Av}{$A_{\rm v}$}
\newcommand{\Rv}{$R_{\rm v}$}
\newcommand{\dg}{$^\circ$}

\begin{document}

   \title{Constraining the structure of the planet-forming region in the disk of the Herbig~Be star HD~100546 \thanks{Based on observations collected at the VLT
   (ESO Paranal, Chile)  with programs 082.D-0010(A), 083.C-0298(A,B),
   083.D-0224(C), 083.C-0146(B), 083.C-0144(A,D), 083.C-0236(A), 075.C-0637(A)}}

\author{E.~Tatulli\inst{1} \and  M. Benisty\inst{2} \and  F. M\'enard\inst{1} \and P. Varni\`ere\inst{3} \and  C. Martin-Za\"{\i}di\inst{1} \and W.-F. Thi\inst{1} \and C. Pinte\inst{1} \and F. Massi\inst{4} \and G. Weigelt\inst{5} \and K.-H. Hofmann\inst{5} \and R. G. Petrov\inst{6}}

\offprints{menard@obs.ujf-grenoble.fr}

\institute{UJF-Grenoble 1 / CNRS-INSU, Institut de Plan\'etologie et d'Astrophysique de Grenoble (IPAG) UMR 5274, 38041 Grenoble Cedex 9, France
\and Max-Planck-Institut f\"ur Astronomie, K\"{o}nigstuhl 17, 69117 Heidelberg, Germany 
\and AstroParticule \& Cosmologie (APC), UMR 7164, Universit\'e Paris Diderot, 10 rue Alice Domon et Leonie Duquet, 75205 Paris Cedex 13, France
\and INAF-Osservatorio  Astrofisico  di  Arcetri,  Largo E.~Fermi 5, 50125 Firenze, Italy
\and Max-Planck-Institut f\"ur Radioastronomie, Auf dem Hugel 69, 53121 Bonn, Germany 
\and Laboratoire H. Fizeau, UMR 6525 UMS, CNRS, OCA. Universit\'e de Nice Sophia-Antipolis, 06108, Nice Cedex 2, France\\}
 \date{Received; Accepted 2011-03-30;}

  \abstract
   {Studying the physical conditions in circumstellar disks is a crucial step toward understanding planet formation and disk evolution. Of particular interest is the case of HD~100546, a Herbig Be star that  presents a gap within the first 13~AU of its protoplanetary disk, a gap that may originate in the dynamical interactions of a forming planet with its hosting disk.} 
   {We seek a more detailed understanding of the structure of the circumstellar environment of HD~100546 and refine our previous disk model that is composed of a tenuous inner disk, a gap and a massive outer disk (see Benisty et al. 2010b). We also investigate whether planetary formation processes can explain the complex density structure observed in the disk.} 
 {We gathered a large amount of new interferometric data using the AMBER~/~VLTI instrument in the H- and K-bands to spatially resolve  the warm inner disk and constrain its structure. Then, combining these measurements with photometric observations, we analyze the circumstellar environment of HD~100546 in the light of a passive disk model based on 3D Monte-Carlo radiative transfer. Finally, we use hydrodynamical simulations of gap formation by planets to predict the radial surface density profile of the disk and test the hypothesis of ongoing planet formation.}
 {The SED (spectral energy distribution) from the UV to the millimeter range, and the NIR (near-infrared) interferometric data are adequately reproduced by our model.  We show that the H- and K-band emissions are coming mostly from the inner edge of the internal dust disk, located near 0.24~AU from the star, i.e., at the dust sublimation radius in our model. At such a short distance, the survival of hot (silicate) dust requires the presence of micron-sized grains, heated at $\sim 1750$K. We directly measure an inclination of $33^{\circ} \pm 11^{\circ}$ and a position angle of $140^{\circ} \pm 16^{\circ}$ for the inner disk. This is similar to the values found for the outer disk ($i \simeq 42^{\circ}$, $PA \simeq 145^{\circ}$), suggesting that both disks may be coplanar. We finally show that 1 to 8 Jupiter mass planets located at $\sim 8$~AU from the star would have enough time to create the gap and the required surface density jump of three orders of magnitude between the inner and outer disk. However, no information on the amount of matter left in the gap is available, which precludes us from setting precise limits on the planet mass, for now. 
 }
{}


   \authorrunning{Tatulli et al.} 
   \titlerunning{The inner disk of HD~100546}

   \maketitle
%

\section{Introduction}
Studying the physical conditions in circumstellar disks and the
processes that rule the evolution of gas and dust provides the context
for the formation of planets. It is now  clear that several processes
are simultaneously acting on the disks at a given time. With the advent of powerful mid- and far-infrared space
telescopes such as \textit{Spitzer}, a new class of objects has been
identified, the pre-transitional and transitional disks \citep[e.g.,][]{espaillat10}.  Pre-transitional disks have 
a typical spectral energy distribution (SED) with a near-infrared excess resulting from the 
emission of hot dust and gas located in an inner disk, a dip in the mid-infrared range likely caused by  a gap, 
and at longer wavelengths, the signature of an optically thick outer disk. 

HD~100546, well studied already by \textit{ISO}, is a late Herbig~Be star (B9.5Ve) that is now classified as a transitional disk.  Located at $\sim$$103^{+7}_{-6}$~pc  \citep[as measured by Hipparcos, ][]{van_den_ancker_2},   it    has   a   large-scale disk inclined by $\sim$40$^\circ$ (0$^\circ$ meaning pole-on, this convention will be used throughout the paper) extending  up  to  4''  that was first imaged in
scattered light \citep{pantin_1}.  Its  estimated age  of  10~Myr \citep{van_den_ancker_1}  corresponds to  the
timescale on  which disks are found to  dissipate \citep{hillenbrand2008}, which suggests that the disk is evolved. Coronagraphic imaging revealed a large-scale envelope ($\sim$1000~AU) and a  disk extending up to 515~AU with
an asymmetric brightness profile \citep{augereau_1,grady_1}.  
Based on its SED that shows a dominant mid-infrared excess, peaking at
40~$\mu$m and with a weak 2-8~$\mu$m emission, \citet{bouwman_1} suggested
the presence of an inner hole within 10~AU that could possibly result
from the presence of a very  low-mass  stellar  or planetary  companion in the disk \citep[20~M$_{\rm{Jup}}$, ][]{acke_1}. 
Nulling   interferometry  at mid-infrared wavelengths showed that the radial temperature law in the
few inner  AUs cannot be explained  with a continuously  flared disk
\citep{liu_1}.   Additional evidence  of  a cavity  is provided  by
means  of  spectroscopy: HST/STIS  observations  of H$_{2}$  emission
revealed a  hole extending to 13~AU \citep{grady_2}, while a  lack of
6300~\AA~[OI]  emission  inside  6.5~AU  is  observed  \citep{acke_1}.
Recent  CO ro-vibrational observations \citep{brittain_1,van_der_plas_1}  reported an
extended  emission  with a  large  inner  hole inside  $\sim$11-13~AU,
confirming the status of \textit{pre-transitional} disk given to HD~100546.

Recent  far-infrared and  millimetric  observations enabled further studies of the gas and dust in the 
outer disk.  \citet{panic_1} detected pure rotational lines of CO and derived a mass of 10$^{-3}$~\msol~  of 
molecular gas in the  disk. Their  model is consistent  with  an  inclined  outer  disk  (i$\sim$40$^\circ$) in
Keplerian    rotation that extends out to 400-500~AU in radius. The \textit{Herschel Space Observatory} detected several additonal molecular lines as well as forbidden atomic lines ([CII] and [OI]) that probably result from photo-dissociation
by stellar photons \citep{sturm2010}. 

On the other hand, very little is known about the inner disk. Far ultraviolet (FUV) observations of molecular 
hydrogen \citep{lecavelier_1,zaidi_1} imply that the observed gas lies very
close to the central star. Although these observations give evidence
that the inner 1.5~AUs are not cleared of gas, clues are still missing
for a firm conclusion about the location of this gas (FUV-driven
photoevaporative wind from the surface layers of the disk, gas in an inner
disk...).  However, the near-infrared (NIR) excess, the emission from
gas located as close to the central star as 0.5~AU \citep{acke_1}, as
well as the signs of accretion \citep{deleuil2004} all suggest that the
innermost AUs are not empty. 

This was confirmed by interferometric measurements in the K-band that spatially resolved the
dust emission at the sub-AU scale \citep[][hereafter   B10]{benisty_1}.  The  authors,   based   on   two
visibility measurements at 2.2 and  8.7~$\mu$m, proposed a  disk model
that  includes an  inner disk,  a gap,  and an  optically  thick outer
disk that are able to reproduce the SED and the visibilities. See their Fig. 3 for a sketch 
of the disk configuration. \citet{benisty_1} considered a stellar luminosity of 22~\lsol~  to reproduce the 
photometric   measurements   at    the   shortest   wavelengths, assuming a null \Av~ and a 
normal \Rv =3.1 extinction law, while other authors previously used a higher value
of  32~\lsol~  using a non-zero \Av~but an anomalous extinction law, \Rv $\sim$5.1 \citep{van_den_ancker_1}.

Owing to the  small number of interferometric measurements used in B10,
some questions remained unsolved.  What is the exact structure of the inner  disk? 
Is it asymmetric and puffed  up, or rather smooth? Is there any hot matter located inside 
the dusty rim that could emit at shorter wavelengths?   Are the inner  and outer disks  coplanar? Could
the presence of a sub-stellar companion be constrained?

In  this paper,  we present  new  spectro-interferometric measurements
obtained  in the  K-band,  as  well as  the  first spatially  resolved
measurements in the H-band and first closure phase measurements for
HD~100546.  We  build  a   complete  SED  from previously  published
photometric and spectroscopic data, and reassess the stellar
luminosity in our model.  Based on the extensive dataset, we revisit the
disk  morphology and  study the effects of a putative
planetary companion on the disk structure and evolution.   In \S~2 we present the  observations and the
data  processing.  In  \S~3 we model the SED and interferomeric data. In \S~4 we use hydrodynamical simulations to predict the surface density profile in the disk when a planet carves a gap into it. \S~5 is the summary.
\begin{table*}[t]
\centering
\caption{Log of the observations and associated $(u,v)$ coverage. All
  data were taken in 2009.} 
\label{tab:obs}
\begin{tabular}{cc}
\begin{minipage}{0.65\textwidth}
\begin{tabular}{cccc||cccc}
   \hline 
\hline 
Date & Baseline & Projected & PA & Date & Baseline & Projected & PA \\
& & length (m)& ($^\circ$) & & & length (m)&  ($^\circ$)\\
 \hline
 2009-01-06 & H0-E0 & 48& 42& 24/04/09 & H0-E0 & 44& -76\\
 & G0-E0 & 16& 42 & & G0-E0 & 15& -76\\
 & H0-G0 & 32& 42 & & H0-G0 & 29& -76\\
 \hline
2009-04-05 & D0-H0 & 62 & 66 &25/04/09 & H0-E0 & 44& -74\\
 & G1-D0 & 61& 119 & & G0-E0 & 15& -74\\
 & G1-H0 & 52& 6 & & H0-G0 & 29& -74\\
\hline
2009-04-08& K0-A0 & 125&72 & 30/04/09 & D0-H0 & 62& 77\\
 & G1-A0 & 85 & 106 && G1-D0 & 62& -55\\
 & K0-G1 & 72 & 32 && G1-H0 & 50& 11\\
 \hline
2009-04-17 & H0-E0 & 46 & 80 & 23/05/09 & A0-D0 & 31& 83\\
 & G0-E0 & 15 & 80&& D0-H0 & 61 & 83\\
 & H0-G0 & 31 & 80 && A0-H0 & 92& 83\\
 \hline 
   \hline 
\end{tabular}
\end{minipage} & \begin{minipage}{0.3\textwidth}
\includegraphics[width=\textwidth]{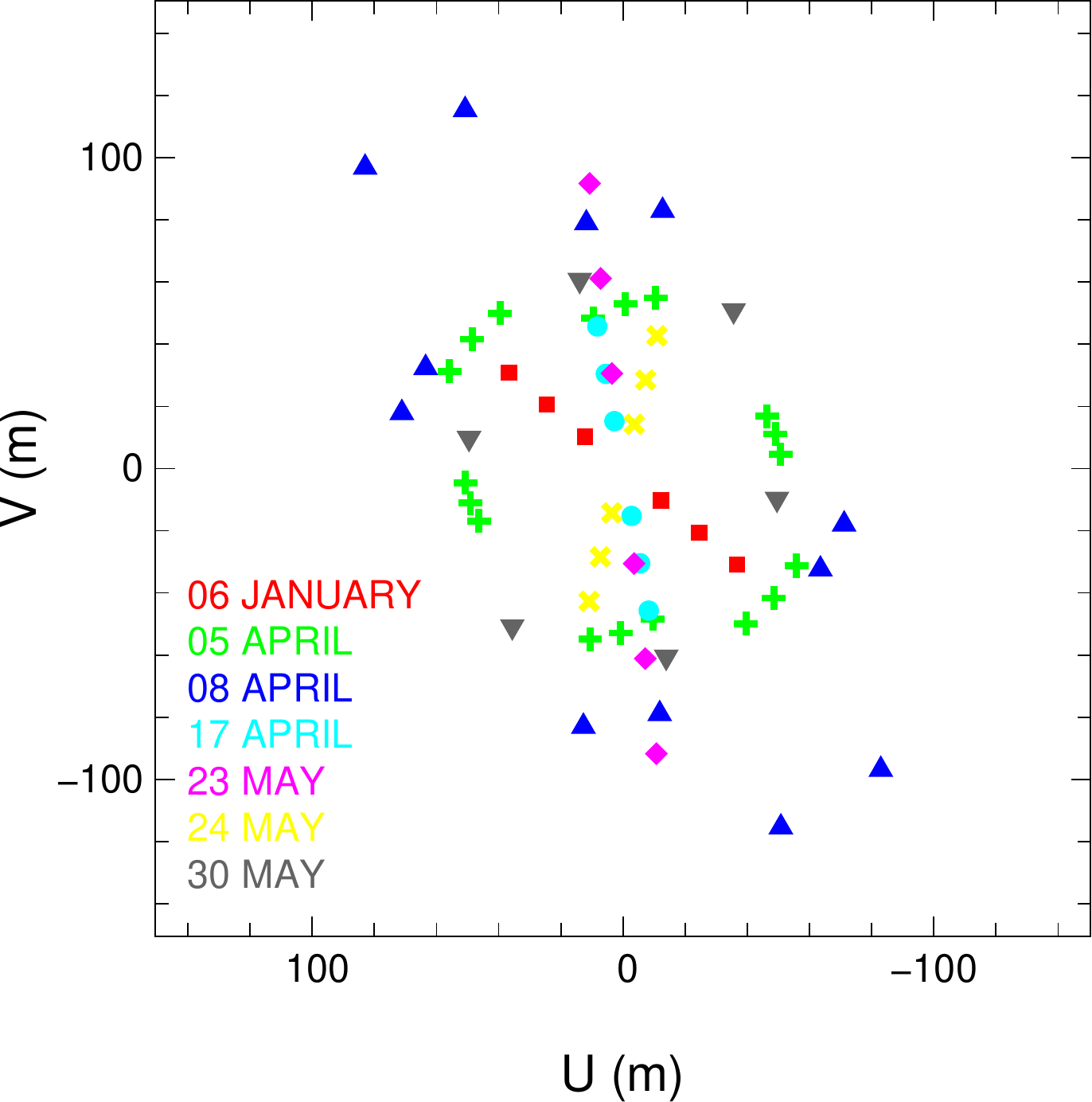}
\end{minipage}
\end{tabular}
\end{table*}

\section{Observations and data processing}
\subsection{Interferometry}
HD~100546 was observed at the Very Large Telescope Interferometer
\citep[VLTI;][]{vlti1}, using the AMBER instrument
that allows the simultaneous combination of three
beams in the near-infrared \citep{petrov07}. It delivers
spectrally   dispersed   interferometric  observables: visibilities,
closure  phases, and differential  phases. In the following, we present H-  and K-band observations  taken in the
low spectral resolution mode  (LR; R$\sim$30) with the 1.8~m Auxiliary
Telescopes (ATs). The data were obtained during several Guaranteed
Time Observation programmes.  HD~100546 was observed with 11 different baselines of
four VLTI configurations during eight nights  from January to May 2009.  The
longest  (projected) baselines achieved for  the H-  and  K-band observations  are $\sim$92 and $\sim$127~m, respectively, corresponding to a maximum angular resolution of $\lambda/B \sim$3.9~mas and 3.6~mas, respectively. We use the
VLTI nomenclature to identify the different configurations.  A summary
of the observations used in this paper is given in Table~\ref{tab:obs}.  
In addition to HD~100546, calibrators (HD104479, HD100901, HD77049) were
observed to  correct for instrumental  effects. Their properties  can be
found in Table~\ref{tab:calparam}.  All observations 
were  performed without the fringe-tracker  FINITO.

The  data  reduction   was  performed  following  standard  procedures
described in \citet{tatulli07} and \citet{chelli09}, using the \texttt{amdlib} package,
release  3.0b4,  and the  \texttt{yorick}  interface  provided by  the
Jean-Marie Mariotti Center\footnotemark{}\footnotetext{$\textrm{http://www.jmmc.fr}$}. 
Raw spectral visibilities and closure phases were extracted for
all the  frames of each observing file.  A  selection of 80\%  of the
highest quality frames  was made to overcome the effects of instrumental
jitter  and  unsatisfactory  light  injection.   Calibration  of  the
AMBER+VLTI instrumental transfer function was done using measurements
of the calibrators, after correcting for their diameter. Finally, calibrated visibilities and closure phases were averaged over the wavelength range of each band (H: from 1.5 to 1.8~$\mu$m; K: from 2.0 to 2.4~$\mu$m), to obtain broad-band observables.
Because of the  combined effects of an intrinsically lower flux
of the source and of a lower instrumental sensitivity, the performance of AMBER is worse in the H-band than in the K-band. As a consequence,  our H-band data are of lower signal-to-noise ratio, which translates into two effects: (i) the
visibility points  show a larger scatter in the H-band, and (ii) we
were unable to estimate  the H-band  closure phase for  the longest
baselines where the source is almost fully resolved. The lower signal-to-noise ratio of the H-band data also explains
the lower number of data points in the lower panel of Fig.~\ref{fig:ringV2}, especially at long baseline lengths where the visibilities are lower.

\begin{table}[t]
\begin{center}
 \caption{Stellar and calibrator properties.}
 \label{tab:calparam}
 \begin{tabular}{c c c c c c}
   \hline 
   \hline 
   Star & K & H & Sp. Type & Diameter  [mas]  \\
   \hline 
   HD100546 & 5.4& 5.9& B9Vne & --- \\
   HD101531 &  4.2 &4.3 & K1III& 0.6$\pm$0.1 \\
   HD104479 & 3.9 &3.9 & K0III& 0.8$\pm$0.1\\
   HD100901& 4.1 & 4.2& K0/K1III& 0.8$\pm$0.1\\
   HD77049& 4.4 & 4.6 & K2III& 0.7$\pm$0.1\\
   \hline 
   \hline 
 \end{tabular}
\end{center}
\end{table}

\subsection{Spectroscopy}
\textit{FUSE spectrum:} Although HD~100546 was observed twice with \fuse, in 2000 and in 2002 \citep{zaidi_1},
we  only  considered  the  exposures  of  2002  that  have  the  best
signal-to-noise ratio. The 
observations were all made in the time-tagged mode, using the
$30''\times30''$ LWRS aperture at a resolution of about 15 000. The
total exposure time was split into several sub-exposures that were co-aligned using a linear cross-correlation procedure and added
segment by segment. The individual spectra (905 - 1187~\AA) were
processed with the CALFUSE pipeline processing
software v3.0.7 \citep{sahnow_1,   dixon_1},   to   correct   for
instrumental effects. 

\textit{UVES spectrum:} The star was observed on 2005
March 21, with UVES \citep{dekker_1}, a cross-dispersed echelle
spectrograph mounted at the VLT.  For these observations, the blue arm was used 
with a spectral coverage between $\sim$3700 \AA\ and 5000 \AA. We used no
image slicer, low gain with a 1$\times$1 binning of the CCDs and a
slit width of $0.4^{"}$  providing the maximum resolution of
$\sim$80,000. Wavelength calibration frames were taken with a long slit
and a ThAr arc lamp. The  spectrum was reduced using the \uves\
pipeline v3.2.0 \citep{ballester_1} available on the ESO Common Pipeline Library.
The spectrum was corrected for the Earth's rotation to shift it in the
heliocentric rest frame. No standard star is available to provide an accurate 
flux calibration. We scaled the UVES flux to the  IUE one, which agrees well with the best photospheric model
for HD~100546 (see Sect. \ref{subsec_stellarprop}). 
  
\textit{IUE spectra:} We  used archived \iue\ spectra at short
(SWP: 1150 - 1975 \AA) and long (LWP: 1850 - 3350 \AA) wavelengths to cover the entire UV band.  
The spectra, of high signal-to-noise ratio,  were obtained  with the large  aperture ($10''
\times  20''$)   and  the  high-dispersion  mode   which  produces  a two-dimensional echelle 
spectrum containing approximately  60 orders, with a resolution of roughly 0.2 \AA. 

\subsection{{\bf The spectral energy distribution}}

The photometric data used in the SED are the same as in B10. They were retrieved from the literature. They include photometry from ground-based telescopes in the optical and NIR,  2MASS data, mid-infrared photometry from the IRAS and  ISO satellites  \citep{ardila_1,henning_1,grady_1, malfait_1, sturm2010}). These are shown as black dots in Fig.~\ref{fig:sed}. We also included spectra from the FUSE, IUE, and ISO archives. 

\section{Modeling}

\subsection{Stellar properties} \label{subsec_stellarprop}
Various estimates of the fundamental parameters of HD~100546
  have been proposed over the past 15 years. The effective temperature
  found in the literature varies from 10\,000 to 11\,000 K, estimates of
  the visual extinction toward this star range from \Av = 0.0
  to \Av = 1.03, and the stellar luminosity estimates range from 22
  to 32~\lsol~ \citep[e.g., ][]{the_1, van_den_ancker_2, valenti_1,  acke_1}. These significant
  differences have motivated a thorough modeling of the 
  photosphere. We started considering photospheric models from the Kurucz grid of LTE models \citep{kurucz_1, castelli_1}. The sub-grid thus obtained samples the parameter space 9000~K $\leq$ \teff\ $\leq$ 12000~K with
250 K steps and 3.5 $\leq$ \logg\ $\leq$ 5.0 with 0.25 dex steps. Then, synthetic spectra were calculated in NLTE using the \synspec\ package of \citet{hubeny_1}. This code allows for an approximate NLTE
treatment of lines with LTE photospheric models\footnotemark{}\footnotetext{$\textrm{http://nova.astro.umd.edu/Synspec43/}$}.
The result of this stage is to determine the LTE model that provides the best fit to the observed spectra. The model assumes a solar
composition, a helium abundance, $Y$ = He/H = 0.1 by number, and a
microturbulent velocity, $\xi _t$ = 2 \kms. For details of the modeling method of FUV and UV spectra 
of Herbig stars, we refer the reader to \citet{bouret_1}.  The best-fit model provides the
following stellar properties: \teff\ = 10500~K, \logg\ = 4.0 and \vsini\ = 55 \kms.

The stellar luminosity was calculated from \teff\ and using HIPPARCOS
parallaxes \citep{van_den_ancker_2, bertout_1} and bolometric corrections using the method described in \citet{bessell_1}:
\begin{equation} 
log~ \bigg( \frac{L_{\star}}{L_{\odot}} \bigg) = \frac {5\times log(1/\pi) -
  m_{\rm v} - BC_{\rm v} -0.26}{2.5} ~~,
\end{equation}
where  $\pi$ is the  parallax  in  arcseconds, m$_{\rm  v}$  is  the V-band
magnitude of the object, and BC$_{\rm v}$, the bolometric correction. We found a luminosity of 25.9$^{+3.5}_{-2.9}$~\lsol.  All  previously published values are within $2\sigma$ of our estimate. In the following, we
use 26~\lsol~. We note that B10 used a stellar luminosity of 22~\lsol~ and a null extinction, assuming a normal (\Rv = 3.1) extinction law. In addition, they had a limited sampling of the stellar flux at wavelengths shorter than 3000 \AA.  A comparison with the stellar models of \citet{siess_1} suggests that this luminosity is probably too low, unless the metallicity of HD 100546 is Z=0.01 (half solar).  

The dataset used by B10 can also be fitted properly with a higher luminosity, e.g., 32~\lsol~as in  \citet{van_den_ancker_1} or with 26~\lsol~(this work) but with a non-zero and anomalous extinction (\Av $\sim 0.2-0.3$ and \Rv $\sim 5.5$). The anomalous extinction law is needed to fit the flat UV spectrum now included in the SED. We note that the photospheric model and extinction curve we use lead to a slight underestimation of the UV flux near 100nm. We suggest this excess is a consequence of the accretion process still going-on for HD~100546. This has no significant impact on the continuum energy budget and continuum emission from the dust disk because the flux densities involved are too low  by a large factor.

\subsection{Location of the near-infrared emission}
\begin{figure}
\centering
\includegraphics[width=0.9\columnwidth]{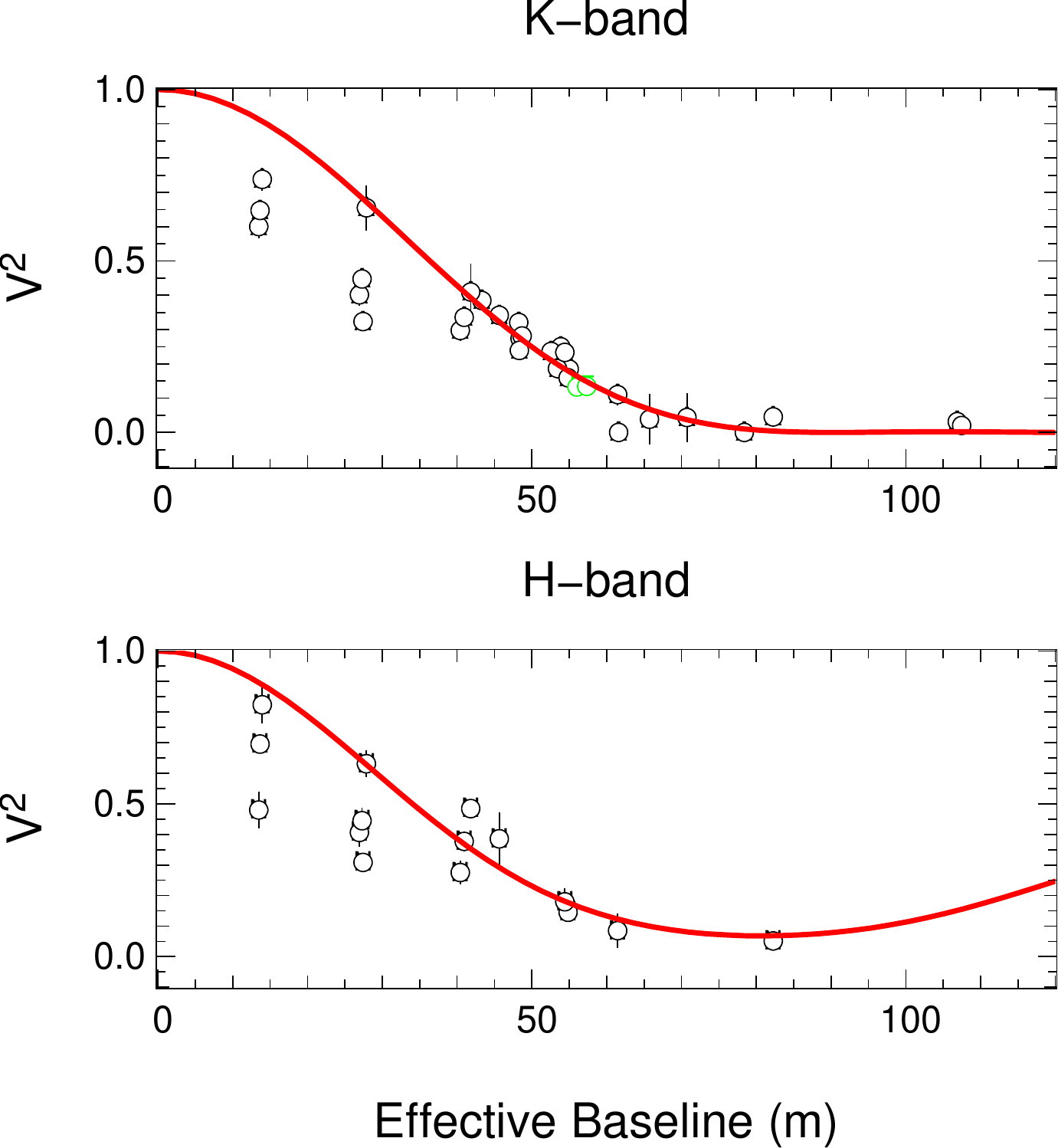}
\caption{Broad-band  squared  visibilities with ($1\sigma$) error bars shown  against  effective
  baseline  for  the  K-band  (upper panel)  and  the  H-band  (lower
  panel). The H-band plot shows less data points because of a lower number of V$^2$ measurements above the required S/N ratio threshold. The  
  predictions  of  the  best  uniform  ring  model  are overplotted in solid line. The filled blue points correspond 
  to the visibilities used in the B10 paper.}  
\label{fig:ringV2}
\end{figure}
To derive the basic characteristics of the NIR emission, we
first  consider  a simple  geometrical  model made of  a ring  of  uniform
surface brightness to roughly  estimate the location
of the emission.  We choose a ring instead of, e.g., a uniform disk because
the bulk of the emission is expected to come from the inner rim of the
dusty  disk, as  shown in  B10. The  ring model  has three  degrees of
freedom, namely the ring inner radius $R_d$, inclination $i_d$ and position angle $PA_d$. The
ring  thickness  is  fixed  at  $20\%$ of  the  radius.  The  modeled
visibility can be written as
\begin{equation}
V^{[H,K]}_{ring} = \frac{V_{\ast}F^{[H,K]}_{\ast} + V_{ex}F^{[H,K]}_{ex}}{F^{[H,K]}_{\ast} + F^{[H,K]}_{ex}},
\end{equation}
 with    $V_{\ast}    =    1$    (unresolved   central    star)    and
 $F^{[H,K]}_{ex}/(F^{[H,K]}_{\ast}+F^{[H,K]}_{ex})$  the  NIR  excess,
 i.e., the excess over the stellar flux divided by the total flux in
 the H- and K-band respectively.  Modelling the stellar photosphere emission as described in Sect. \ref{subsec_stellarprop},  we estimate  H- and  K-band excesses of $\simeq 0.53$ and $\simeq 0.75$.
  
 An independent fit of the H- and K-band visibilities for HD~100546 shows that the continuum emission in both bands 
 comes from the same location, within the uncertainties. We therefore performed a classical
 $\chi^2$  minimization using the complete  H- and  K-band visibility datasets together. 
 We constrain the NIR-emitting region to be located at $R_d = 0.25$~AU $\pm~0.02$~AU, which reinforces 
 the estimation made in our previous study ($R_d = 0.26$~AU;  B10).  We note that this estimation is independent of the continuum emission mechanism, it is in particular independent of the dust properties in the disk.
  
 Because the observations provide a good coverage of the spatial frequencies (see Table~\ref{tab:obs}) we are 
 also able to estimate the inclination and position angle of the inner disk. They are $i_d = 33^{\circ} \pm~
 11^{\circ}$ and $PA_d = 140^{\circ} \pm~ 16^{\circ}$ respectively. The best ring model is  shown   in  Fig.~\ref{fig:ringV2},  together  
  with   the  squared visibilities  plotted against the effective baseline length. The effective
 baseline length is defined as
\begin{equation}
B_{\rm{eff}} =  B\sqrt{cos^{2}(\theta) +  cos^{2}(i_d)sin^{2}(\theta)},
\end{equation} 
where $\theta$ is the angle between the baseline direction and the major
axis of  the disk, and $i_d$  is the disk  inclination \citep{tannirkulam}.  This
representation is useful to show a large dataset in a concise way once the
inclination and position angles of the disk are known. 
Our estimates, relevant for the inner disk, agree well with the previously published
  values of both the inclination and position angles of the disk measured at larger scales \citep[e.g., ][]{ardila_1}.
This result suggests  that  the inner  and outer disks of HD~100546 are most likely  coplanar -- as  implicitly assumed in B10
 -- or very close to it. This is unlike the case of HD~135344~B for example, another transitional disk \citep[][]{grady09}. In the rest of this paper we will use a single inclination and position angle for the  whole  disk, adopting  the
 values commonly  used in previous modeling,  namely $42^{\circ}$ and
 $145^{\circ}$, respectively.
  
We remark that if our fit is good for the long ($\ga 30$m) effective baselines (i.e., for angular resolution better than $\sim$ 12~mas),  our simple ring model is systematically too high with respect to the visibilities at short baselines. Although most
 of the  NIR emission is located around $0.25$~AU,  this may indicate that  there is some extra extended emission not included  in the  ring  model. The  origin of  this extended emission  will be  discussed in the next section,  together  with  our radiative  transfer  disk  model that describes the disk environment of HD~100546.  

\begin{figure}
\centering
\includegraphics[width=0.9\columnwidth]{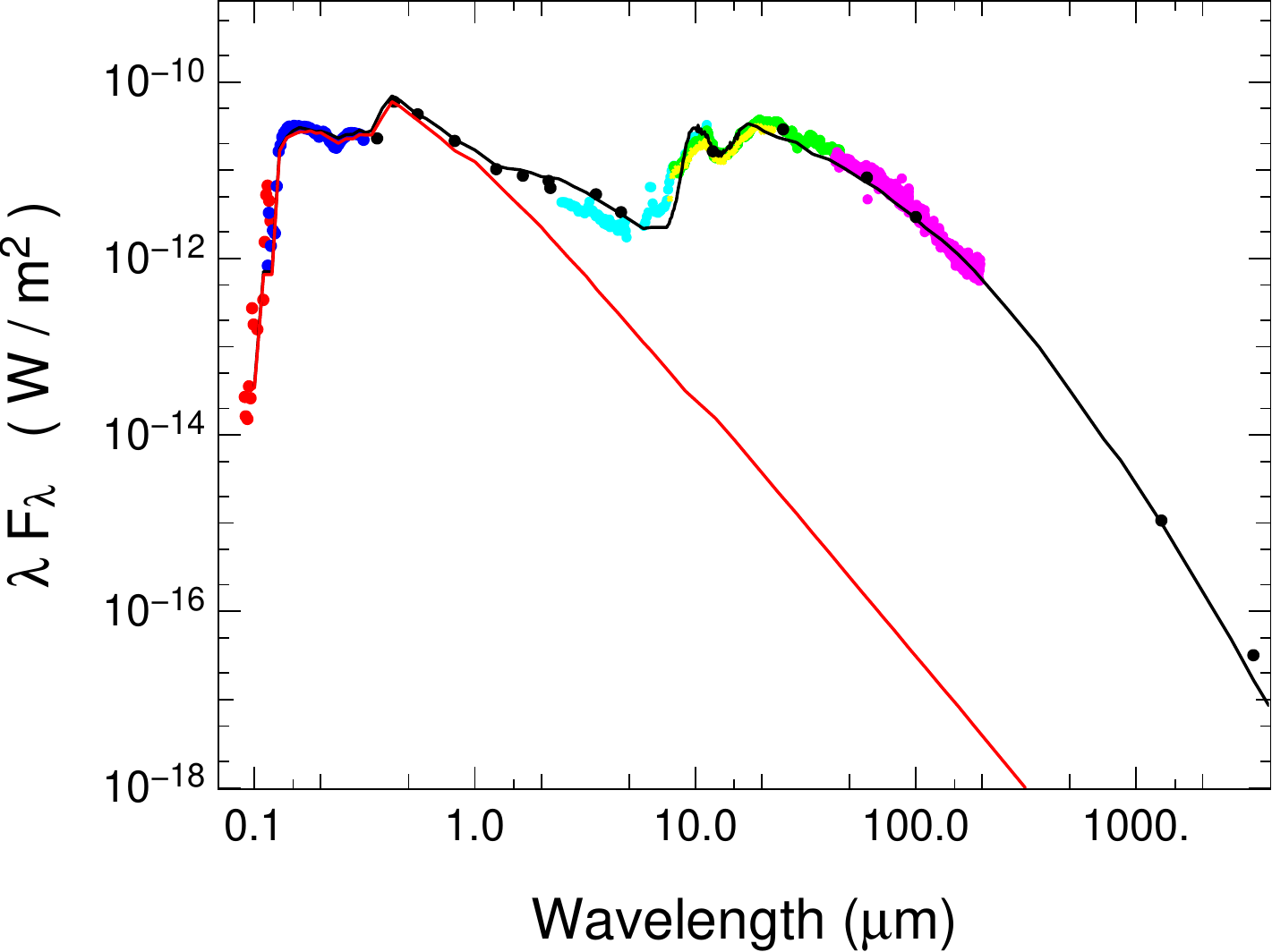}
\caption{Observed spectral energy distribution of HD~100546 compared to our
  best-model prediction (full black line, Teff=10500~K, 26~\lsol, \Av=0.2, \Rv=5.5). The photosphere is shown as a red line.
  The photometric data used in B10 are represented by black dots. The binned FUSE spectrum is plotted with red dots, the binned IUE spectrum with blue dots, the archival ISO data are plotted with cyan dots for ISO-PHOT, green dots for ISO-SWS, and magenta dots for ISO-LWS. The IRAS-LRS data are represented by yellow dots.}
\label{fig:sed}
\end{figure}
\subsection{Disk structure}
\begin{table}[tdp]
\caption{Best-model parameters.   Boldface indicates  when  the value
  differs from the B10 model.}
\begin{center}
\begin{tabular}{l l || c c c }
\hline
\hline
 & & inner disk & surface layer & outer disk \\
\hline
$M_{\rm{dust}}$ & [\msol] & $\vec{1.75\times 10^{-10}}$ & $\vec{3\times 10^{-7}}$ & $\vec{4.3\times 10^{-4}}$\\
$R_{\rm{in}}$ & [AU] & {\bf 0.24} & 13 & 13 \\
$R_{\rm{out}}$ & [AU] & 4 & 50 & {\bf 500}\\
$H_{\rm{100AU}}$ & [AU] & 6 & 12 & 12\\
 $\beta$ & & 1 & 0.5 & 1.125\\
$p$ & & -1 & -1 & -1\\ 
$a_{\rm{min}}$ & [$\mu$m] & 0.1 & 0.05 & 1\\
$a_{\rm{max}}$ & [$\mu$m] & 5 & 1 & 10000\\
\hline
\hline
\end{tabular}
\end{center}
\label{tab:modelparam}
Note: $\beta$ is the exponent of the scale height radial profile, $p$ is the exponent of the surface density radial profile.The grain size distribution is a power-law with a slope of -3.5.
\end{table}
{\it Radiative transfer with MCFOST:} Going beyond  the simple geometrical
ring model, we used the  Monte Carlo-based 3-D radiative transfer code
MCFOST  \citep{pinte_1, pinte_2}   to  compute  the  SED  and   NIR  images  of
HD~100546. Visibilities  and closure phases are computed  from the NIR
model images. Here, we refine the model developed in B10  that was based on
one  visibility  measurement. In particular, we use a higher stellar
luminosity of $26$~\lsol~(see, Sect.~\ref{subsec_stellarprop}). 

The disk  is similarly  composed of  (i) an inner  dusty disk  with an
inner edge at 0.24~AU, (ii) a gap between 4~AU and 13~AU, necessary
to reproduce the dip in the SED, (iii) a surface layer (over the outer disk) of small grains, and (iv) an outer disk from 13~AU to 500~AU holding most of the  mass  (see the sketch of the disk in Fig.~3 of B10). It is difficult to provide reliable error bars for all these values without a full exploration of the parameter space. But the uncertainties, or validity ranges, in particular regarding the position and size of the gap, were discussed in B10 (see \S5). Their discussion is still valid. 

Considering our larger  dataset and the change in the stellar  luminosity,  some of  the  B10  model  parameters need  to  be
slightly adjusted. The location of the inner rim is slightly decreased from  0.26  to  0.24~AU to better match the visibilities. 
We also reduced the mass of both the  inner disk and of the surface layer (see Table \ref{tab:modelparam}) to reproduce the NIR excess as well as the MIR bump in the SED.  The outer disk now extends up  to 500~AU, instead of 350~AU, following recent CO  line observations by  \citet{panic_1}. We modified the  outer disk mass accordingly. In order to generate the large silicate emission feature seen in the SED at $10\mu$m, we used amorphous olivine grains \citep{dorschner_1} for the composition of the surface layer, whereas the rest of the disk is made of astronomical silicate grains \citep{draine_1}, as in B10. The remaining  parameters such as  density profile, scale height, flaring and grain sizes were kept unchanged. The parameters are summarized in
Table~\ref{tab:modelparam}. This model  provides an adequate fit of the SED from the UV to the millimeter range, as shown in Fig.~\ref{fig:sed}.
\begin{figure*}[t]
\centering
\includegraphics[width=\textwidth]{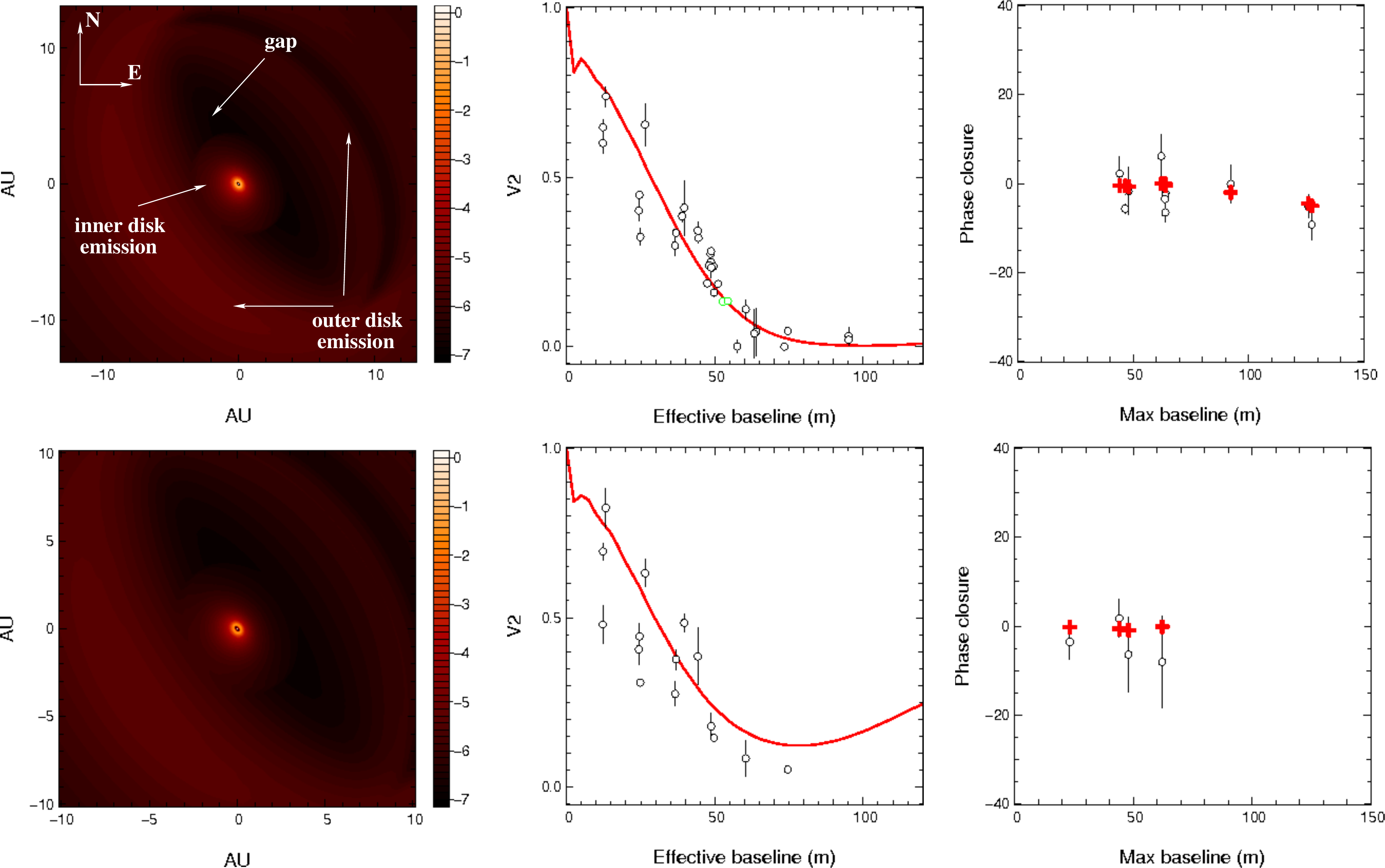}
\caption{K-band (top) and H-band (bottom) MCFOST modeling of HD~100546. From left to right: NIR images (normalized intensity to one at maximum, in logarithmic scale), visibility (red solid lines) and closure phase (red crosses),  compared with the interferometric observations (black circles and error bars). In the middle panels, the "kink" in the model visibility curves at  B$\sim$10~m is a real feature caused by the sharp inner edge of the outer disk.}
\label{fig:modelV2}
\end{figure*}

Fig.~\ref{fig:modelV2} compares the model predictions with the observations.  The models agree well with the entire set of H- and K-band  visibilities  and   closure  phases.  The "kink" in the model visibility curves at B$\sim$10m is caused by the outer disk that has a sharp inner edge and with a finite sampling of the baselines in the models. Increasing the sampling makes the curve smoother, but the "kink" remains. A further demonstration of this  can be seen in Fig. 4 (top row, middle panel), where the "kink" disappears when only the inner disk is considered in the field-of-view of the interferometer.
 
 In addition to the visibilities that probe the inner rim location and the properties of the dust it contains, we now 
discuss the first closure phase measurements  of  HD~100546. The closure phase is related to
the  level  of  asymmetry of the source, here of the NIR continuum emission.
Roughly, the values of the closure  phases, given in radians,  give the ratio
between the asymmetric and symmetric fluxes for the corresponding
telescope triplet \citep{monnier_1}. As theoretically 
expected  \citep{lachaume_1},  the departure from centro-symmetry becomes
significant for the closure phase for the longest baselines only, i.e., when the source  is well
resolved. In our case, with $B_{max} = 127$~m we find a closure phase signal of  $\sim 5-10$\dg~in the K-band, whereas it is compatible with $0$\dg~in the H-Band. Recall, however, that the 
uncertainties are larger and there is a lack of visibilities at very long baselines at H.
\begin{figure*}
\centering
\includegraphics[width=\textwidth]{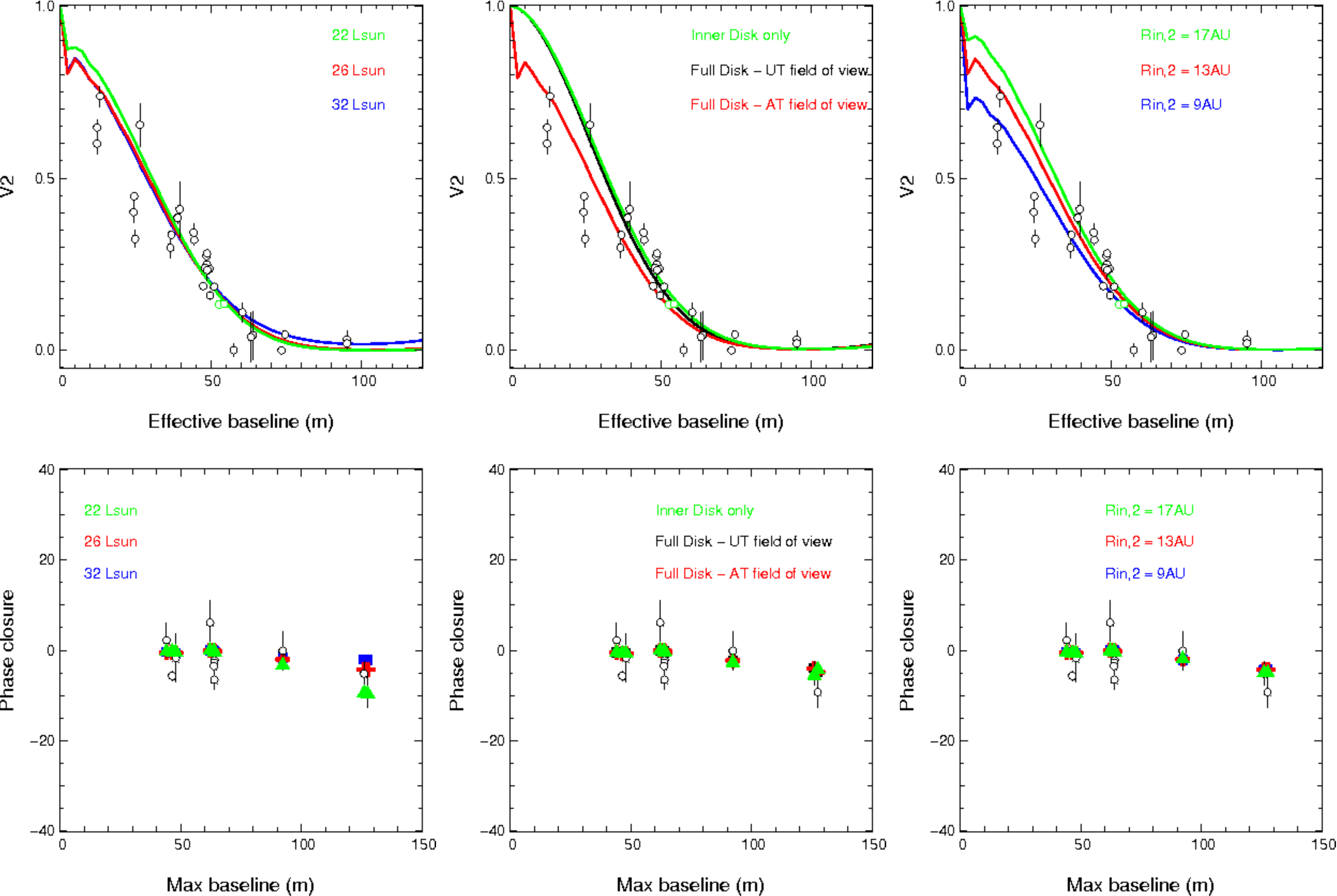}
\caption{K-band MCFOST modeling of visibility (top) and closure phase in degrees (bottom) for  various values  of model parameters.  Left: Various  stellar   luminosities  are   compared:  22~\lsol~(green),
  26~\lsol~(red), and  32~\lsol~(blue). Middle: Effect  of the field-of-view when  
  using  the  AT (red) or  UT  (black) telescopes.  With  the latter,  only the inner disk of  HD~100546 is
  seen, which is shown in  green. The black and green curves lie on top of each other, as expected. Right: the interferometric observables are
  plotted for different locations of the inner edge of the outer disk,  $R_{in,2}=$  9~AU  (blue),  13~AU  (red), and  17~AU  (green).} 
\label{fig:compare}
\end{figure*}

{\it  Properties  of  the  inner   disk:}  From the data we constrain the location of the disk inner edge
at $r \sim 0.24$~AU from the star. Assuming silicate grains, the inner disk is composed of micron-size particles whose 
albedos make the scattered light an important part  ($\sim 30\%$) of the disk total  K-band  emission, as
discussed in B10.  Fairly large grains need to be used (i.e., with $a_{\rm max} = 5~\mu$m)  to make them survive (i.e., not sublimate too fast) at the inner edge of 0.24~AU, a value imposed by the interferometric observations. Because of the higher stellar luminosity than the one used in B10, we find a temperature  for the hottest grains of  $\sim 1750$~K. We note that this value is on the warm side for silicate grains but we did not attempt to use different compositions.  We did not explore either the possiblity that much larger grains be present in the inner disk because their contribution to the disk emission in the NIR is not dominant with respect to the grains used here. Because of the small NIR excess in the SED, we find that the inner disk must be tenuous, with a dust mass of $\simeq$ 1.75$\times 10^{-10}$~\msol. 
 
We find an integrated radial optical depth of $\tau \simeq 10$ (in the I-band) in the equatorial plane of the inner disk. Because it is opaque, depending on its geometrical thickness it may (or may not) cast a shadow and may (or not) significantly modify the temperature structure of the outer disk, hence its scale height. We verified that the scale height profiles used in our parametric models agree with a disk in hydrostatic equilibrium. To check this, we calculated the hydrostatic scale height of a vertically isothermal disk, consistent with the vertical Gaussian density profiles we used, and used the two extreme temperatures found at each radii to bound the  scale height of the actual disk. As a minimum temperature for each radius  we used the temperature of the disk midplane, the maximum temperature being estimated at the disk surface. We find that the scale height of the outer disk (i.e., H=12~AU at r=100~AU) falls between the two estimates, which suggests that the inner disk does not dramatically modify the structure of the outer disk. We note that the scale height of the inner disk (6~AU at 100~AU) is slightly too large, by about 30\%.

The measured closure phase value of $\sim 5-10$\dg~ in the K-band is  well  reproduced  by  our  model.  It is a direct consequence of the {\it anisotropic scattering of starlight by the dust in the inner disk}. The phase function of the dust grains in our model has an asymmetry factor $g \sim 0.6$ ($g=0$ is isotropic scattering and $g=1$ is fully forward-scattering). The forward-throwing nature of the light scattering process in our model directly results in the  asymmetry of the NIR brightness distribution because the disk is inclined.  Note also that the closure  phase signal is  entirely caused by the inner disk, because the outer disk  is too faint to contribute at K-band, as shown in Fig.~\ref{fig:compare} (middle). The outer disk has a different
impact on the visibilities, see below. We remark here that we  do not need to  invoke the presence of  a ``puffed-up'' inner
rim  \citep[e.g., ][]{dullemond_1,  isella_1} to  reproduce  the
interferometric data because the scale height of the inner rim follows the smooth law defined for the whole inner disk, i.e.,  $H(r) = H_0(r/r_0)^{\beta}$. In our model we thus have $H(R_{\rm{in}}) = 0.015$~AU corresponding  to a ratio H/R of $0.06$ for the inner rim, this is a factor 2 to 4 smaller than the values of H/R$\sim 0.1 - 0.25$ that are typically expected when the rim is puffed-up and vertical \citep{dullemond_1}.  \citet{monnier_2}  showed   that  sharp ``puffed-up''  inner  rims  are
expected to produce large closure phase signals as soon as they
are  resolved  enough. In our  case, the value of the  inner rim diameter ($0.48$~AU)
in  units of fringe  spacing $\lambda/B_{max}$  is $\sim  1.3$, a  value for
which   substantial   closure   phases  of $\ga   40$\dg~  would be expected
\citep[see][Fig.~22]{monnier_2}. Such high values are clearly ruled out by our
data.  A sharp ``puffed-up''  inner rim  for HD~100456  is unlikely.  

 The closure phase signal is also very sensitive to the value
of  the flux  ratio  between the  star  and the  disk.  Because the ratio  will  vary   with  the   stellar  luminosity,   a  careful
determination of  this parameter  is essential. The study of the closure phase is a good way to cross-check our luminosity estimation of Sect. \ref{subsec_stellarprop}.
Figure~\ref{fig:compare} (left) shows  the behavior of the (visibility
and)  closure  phase  for  L$_{\star}  =  22$~\lsol~,  $26$~\lsol,  and
$32$~\lsol,  including   the  values  used  in   the  literature. While the visibility profile 
does not  change significantly with the luminosity,  the closure phase
increases  from a  few  degrees  to $15$\dg~  as  the stellar  luminosity
decreases. Such analysis indirectly  confirms that the  luminosity of
HD~100546  must indeed lie in the  [$22$,$32$]~\lsol~ range. More
accurate  data would  be of  great interest  to reduce  this  range of
validity.

{\it Gap and  outer disk:} The Auxiliary Telescope's field of  view of $\lambda/D \simeq
250$~mas $\equiv 25$~AU enables us to probe the outer environment
of  HD~100546, that  is,  the gap  and  the outer  disk. This  extended
emission within the  field of view directly impacts  on the visibility
profile at short baselines, as shown in Fig.~\ref{fig:compare} (middle
and right). Including the emission from the outer disk in the model and the field of view better reproduces the visibilities at short-baselines -- despite their scatter --  than the ring model  alone.  Typically, with short
baselines  of a  few tens  of meters, the interferometer becomes mostly sensitive  to the
location  of  the inner  edge  of the  outer  disk,  as emphasized  by
Fig.~\ref{fig:compare}  (right). This suggests that  the  gap must end
somewhere between 9~AU and 17~AU. Unfortunately, the scatter
in the data does not allow for a better estimation of this parameter. \\

Interestingly enough, the  outer part of the disk  would not have been
traced with  the $8$~m Unit Telescopes of the VLTI because it lies outside its
field-of-view ($\lambda/D \simeq 60~$mas $\equiv 6$~AU). This points out
the  importance of  the auxiliary  telescopes,  not only  in terms  of
adding baseline  capabilities to that of  the UTs, but also  in terms of
field-of-view for tracing potential extended emission.  Indeed, lower visibilities at short baselines with respect to the uniform ring case were also measured for other Herbig AeBe stars \citep{monnier_2} and were interpreted as resulting from an extended halo. Our model shows that the scattered light emission from the outer part of a flared disk (even when continuous), a contribution that is usually neglected when modeling the K-band emission, can naturally and easily explain this fairly common feature. 

\section{Hydrodynamical simulations of gap formation and inner disk evolution}
\begin{figure*} %
   \centering
\begin{tabular}{cc} 
  \includegraphics[width=0.45\textwidth]{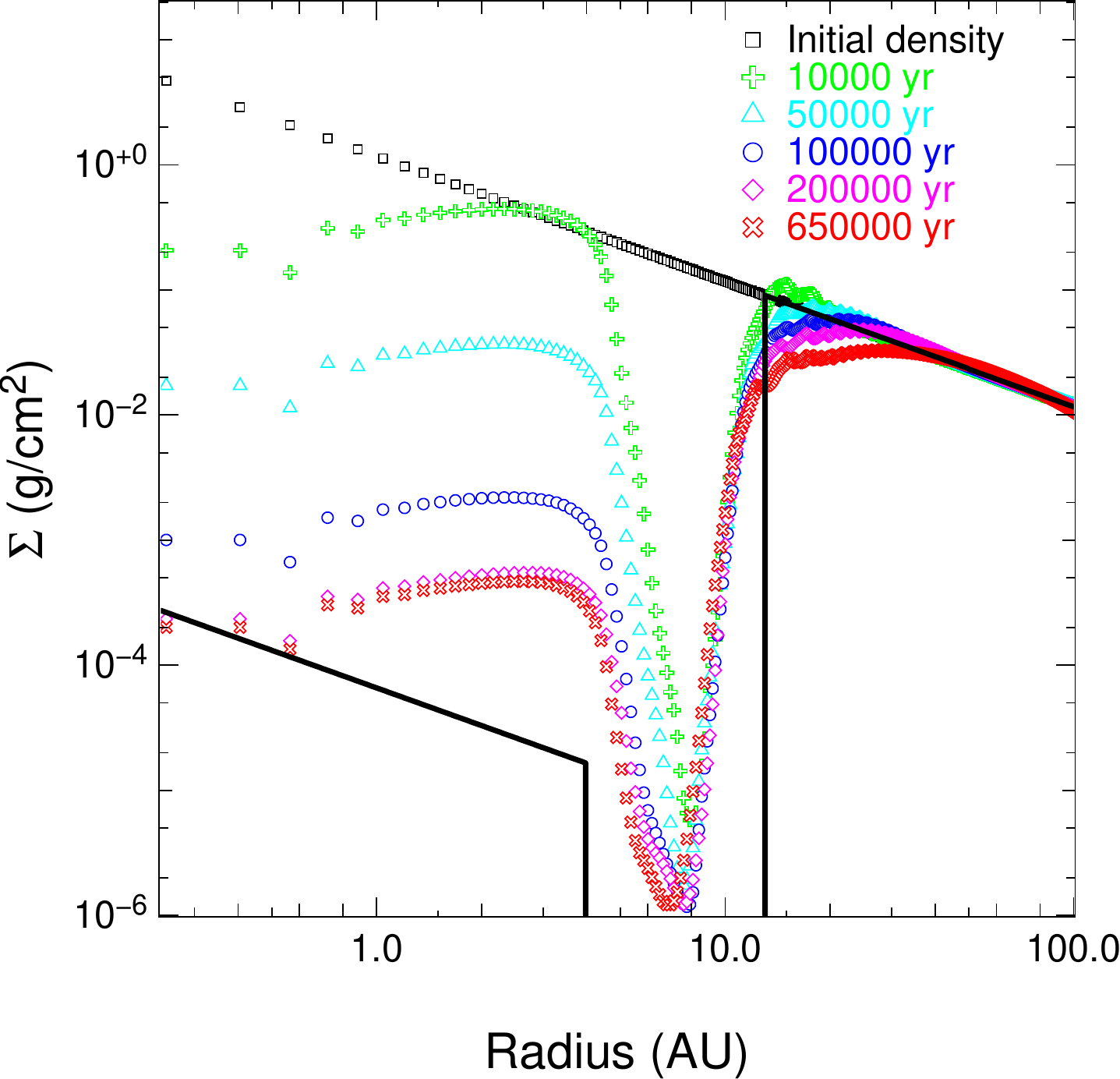}&
 \includegraphics[width=0.45\textwidth]{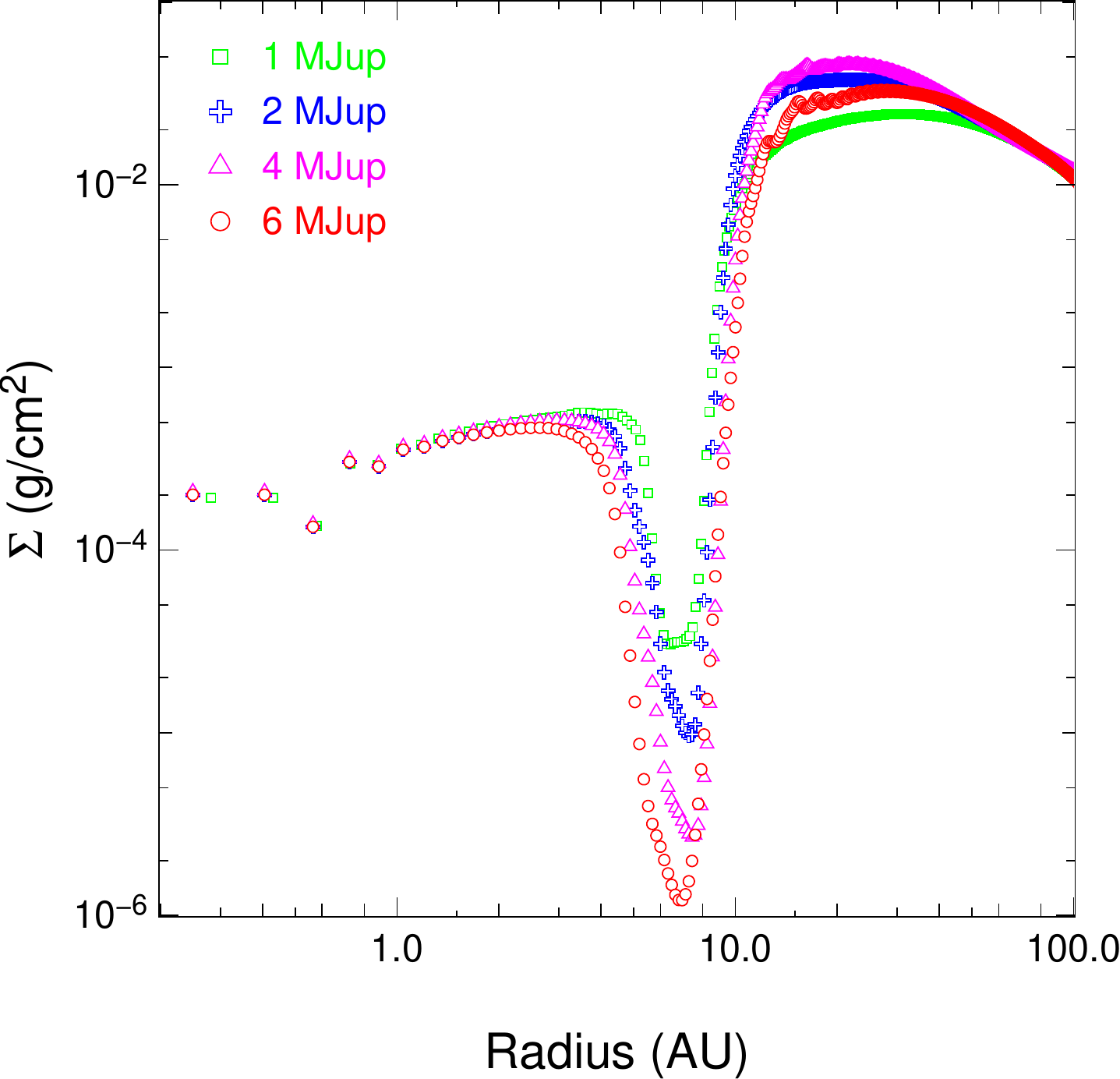}
\end{tabular}
 \caption{Left: evolution of the density as function of radius in the
case of a 6 Jupiter-mass planet. The solid line indicates the density profile used in our model. Right: final density profile as a function of the disk radius for various planet masses.}
   \label{fig:inner_disk}
\end{figure*}

It is well documented that a planet in a disk is expected to carve a gap, a trench, in its density profile.
Numerous studies exist to study the formation of these gaps by planets via numerical simulations
\citep{bryden_1,kley_1,nelson_1,V04}, but none to our knowledge has focused on trying to reproduce the density jump values needed to match the observations.  In this paper, our goal is not to revisit the gap formation mechanism but rather to verify whether a planet embedded in the disk of HD~100546 can explain the large jump in the surface density profile observed on both sides of the gap. To do so, we carried out a series of  2D simulations involving planets of various masses and opening disk gaps. We have made several assumptions and simplifications: circular orbit for the planet, constant accretion rate onto the star over the its full lifetime and equal to the present day accretion rate, the accretion rate in the disk is the same as onto the star and the same at all radii in the disk.  However, we have run the numerical simulations sufficiently long to let the inner disk evolve until steady-state is reached. Following \citet{acke_1}, the lowest planet mass estimate based on the local disk thickness is about 1 Jupiter mass, we therefore started our study with a one Jupiter-mass planet.

To reproduce the density jumps, several mechanisms are competing in the inner region. These are (i) The accretion onto the star; for HD$100 546$ we have an estimated accretion rate of $\sim 10^{-9}$ \msol/yr \citep{deleuil2004, grady_3}. This will tend to deplete
the inner region, which in turn needs to be replenished. Indeed, with the estimated mass available in the inner region and
the observed accretion rate, the inner region should empty itself in less than one year. (ii) The planet "pushes" the gas in the inner region during the creation of its gap. This is a transitory event and occurs soon after the planet formation, it is not part of the steady state we are looking for. (iii) Some material from the outer disk can flow accross the gap, cross the planet's orbit, and replenish the inner region.

To reproduce the density jump suggested by the observations of HD~100546,
one needs to reproduce a surface density for the inner disk that is about three orders of magnitude lower than the surface density just outside of the gap. Ideally, it also needs to produce the proper gap width, i.e., between about $4$ and $13$~AU.

\subsection{Hydrodynamic code and simulation setup}
To cover  a sufficiently long time-span in the simulation we used the algorithm {\tt
FARGO} \citep{m00,m02},  which eliminates the azimuthal velocity from the computation of the
 Courant-Friedrich-L\'evy condition. This speeds up the computation
and  facilitates the study of the long-term disk evolution. Then, to follow the disk over several orders of magnitude in radius to see how the density evolves close to the star as well as outside, we used the code {\tt FARGO-2D1D} \citep{C07}, which is an
extension of the standard version of {\tt FARGO}, where the 2D grid is surrounded by a simplified 1D grid made of
elementary rings that are non azimuthally resolved. It aims at studying and taking into account the general evolution of
the disk in the simulations of planet-disk interactions.

We ran several simulations to find and optimize the planet mass
needed to create the required density jump. A planet will open a gap  near its Lindblad resonance located at $r_L
= \left( 1  + \frac{1}{m}\right)^{2/3} r_p (1\pm e)$ where $m$ is the mode of the Lindblad resonance, $r_p$ the 
position of the planet and $e$ the eccentricity of the orbit. To obtain the outer edge at $\sim 13$~AU, we placed the planet at $8$~AU. We set up the 2D disk around it to capture the 2D structure of the
wave. The 2D disk runs from $4$ to $20$~AU and the 1D
disk extends from $0.1$~AU to $150$~AU\footnote{We ran test
simulations with different sizes of both disks to validate the range of the 2D
vs 1D part of the domain and found no major differences.} to reproduce
 the global behavior of the whole region.

\subsection{Simulated density profile: Evolution with time}
In Fig.~\ref{fig:inner_disk} (left) we follow the time evolution of
the surface density in the first $100$~AU of the disk for a $6$ Jupiter-mass planet. 
Such a planet produces a surface density jump of the correct amplitude: the inner disk surface density
drops by three orders of magnitude in about $200\ 000$ years, confirming that a 6~M$_{Jup}$ planet located at 8~AU can carve the disk of HD 100546 with the proper surface density jump across the gap.

Figure \ref{fig:inner_disk} (right) shows that most planet masses are able to reproduce the strong density jumps imposed by the observations. The only differences are the time required to reach that state, the amount of mass left in the gap, and the width of the gap. The main parameters that regulate how much mass is left in the inner region are the accretion rate onto the star and the disk
viscosity. If we can estimate the accretion rate from observation, it is a lot
harder to have access to the local disk viscosity. We thus adopted an {\it ad hoc} value of the viscosity that was successfully used for previous simulations \citep[see for example][]{quillen_1} and aim only at reproducing the density jump. 
This does not allow us to provide a lower limit for the mass of the putative planet because the longest timescale is well within the age of the system and so far we do not have constraints on the mass inside the gap.\\

Further remarks can be made regarding these numerical results. It is interesting to note that the density profile of the inner region becomes flat early on (before $10\ 000$ years) owing to the accretion onto the star. It becomes flatter than the $r^{-1}$ power-law used in our model. However, as mentioned above, the slope of the density profile is not well constrained by the current observations. We verified that radiative transfer models that use an inner disk with a flat inner density profiles (i.e., constant with radius) also match the interferometric data and SED.\\

Also, during the  $700\ 000$ years of the simulation time span the planet has a slow migration from $8$ to about $7$~AU. This slightly moves the outer edge of the gap but it remains within the observational limit.  However, in all simulations the gap width appears too narrow compared to the surface density profile estimated from the data. We  did not try to solve this question.
We can only suggest that an elliptical orbit for the planet may help to broaden the gap. Also, HD~100546 is significantly older than the final time of each simulations. It is therefore difficult to estimate how much wider the gap would have grown in the simulations by the age of the star. Finally, we also note that the full extent of the inner disk, in particular its outer radius, are not well contrained by the data. R$_{out}$ of the inner disk, which defines the inner boundary of the gap, could be moved outward somewhat depending on the exact vertical density profile. We refer the reader to a forthcoming paper for a more detailed comparison of the profiles from the hydrodynamical simulations and those from data fitting. For now, we only  note that a planet can produce the observed surface density jump across the gap.

\section{Summary}
The main results of our paper are summarized below:
\begin{itemize}
\item using the AMBER/VLTI interferometer in H- and K-band, we spatially resolved the near-infrared emission region of HD~100546. Most of this emission arises from the inner edge of its inner disk located at $0.24\pm0.02$~AU from the star 
\item  a puffed-up rim in the inner disk appears unlikely because of the small closure phase signal
\item for the first time, our observations also constrain the inclination and the position angle of its inner disk. We find $i= 33^{\circ} \pm
 11^{\circ}$ and $PA = 140^{\circ} \pm 16^{\circ}$, which suggests that the inner and outer disks are likely coplanar.
\item  we described the circumstellar environment of HD~100546 in the light of a passive disk model based on 3D Monte-Carlo radiative transfer. Our model is composed of (i) a tenuous inner disk with a dust mass of  $\sim$1.75$\times 10^{-10}$~\msol~ from $0.24$~AU to $\sim$4~AU, (ii) a gap devoid of dust, and (iii) a massive outer disk with a dust mass of $\sim$4.3$\times 10^{-4}$~\msol~ extending from $\sim$13~AU to $\sim$500~AU. The model reproduces both the SED and the interferometric observations adequately.
\item we show from hydrodynamical simulations that disk-planet interactions with a planet located at $\sim$8~AU are able to reproduce the observed density jump between the inner and outer disks, across the gap.
\end{itemize}

\begin{acknowledgements}
We thank the VLTI team at Paranal.  C. Pinte acknowledges funding from the European Commission seventh Framework Program (contracts PIEF-GA-2008-220891 and PERG06-GA-2009-256513). F. M\'enard, C. Pinte, C. Martin-Za\"{\i}di and W.-F. Thi acknowledge PNPS, CNES and ANR (contract ANR-07-BLAN-0221) for financial support.
\end{acknowledgements}

\bibliographystyle{aa}
\bibliography{hd100_myriam}

\end{document}